\documentclass[num-refs]{nbdt-article}

\usepackage{siunitx}

\usepackage{ccfonts}

\usepackage{graphicx}
\usepackage{longtable}
\usepackage{booktabs}
\usepackage{color}

\papertype{Original Article}

\pdfoutput=1

\title{Beyond linear regression: mapping models in cognitive neuroscience should align with research goals}

\author[1,2
]{Anna A. Ivanova}
\author[1,2,5]{Martin Schrimpf}
\author[3]{Stefano Anzellotti}
\author[1,2,5]{Noga Zaslavsky}
\author[1,2]{Evelina Fedorenko}
\author[4]{Leyla Isik}

\affil[1]{Department of Brain and Cognitive Sciences, MIT}
\affil[2]{McGovern Institute for Brain Research, MIT}
\affil[3]{Department of Psychology, Boston College}
\affil[4]{Department of Cognitive Science, Johns Hopkins University}
\affil[5]{Center for Brains, Minds and Machines, MIT}

\corraddress{Anna A. Ivanova}
\corremail{annaiv@mit.edu}

\fundinginfo{AI was supported by the Whitaker Health Sciences Fund Fellowship. MS was supported by Fellowships from Takeda and the McGovern Institute, and the SRC Semiconductor Research Corporation. NZ was supported by a BCS Fellowship in Computation and a K.~Lisa Yang Integrative Computational Neuroscience (ICoN) Postdoctoral Fellowship. EF was supported by R01 awards DC016607 and DC016950, a U01 award NS121471, and research funds from MIT BCS, the McGovern Institute for Brain Research, and the Simons Center for the Social Brain.
}

\runningauthor{Ivanova et al.}

\begin{document}

\pagecolor{white} 
\maketitle

\begin{abstract}

{Many cognitive neuroscience studies use large feature sets to predict
and interpret brain activity patterns. Feature sets take many forms,
from human stimulus annotations to representations in deep neural
networks. Of crucial importance in all these studies is the mapping
model, which defines the space of possible relationships between
features and neural data. Until recently, most encoding and decoding
studies have used }{linear }{mapping models. Increasing availability of
large datasets and computing resources has recently allowed some
researchers to employ more flexible }{nonlinear }{mapping models
instead; however, the question of whether nonlinear mapping models can
yield meaningful scientific insights remains debated. }{Here, we discuss
the }{choice of a }{mapping model in the context of three overarching
}{desiderata:}{~predictive accuracy, interpretability, and }{biological plausibility}{. }{We show that, contrary to popular intuition, these }{desiderata}{~do not map cleanly on}{to the linear/nonlinear divide}{;
instead, each desideratum can refer to multiple research goals, each of
which imposes its own constraints on the mapping model. Moreover, we
argue that, instead of categorically treating the mapping models as
linear or nonlinear, we should instead aim to estimate the complexity of
these models. We show that, in many cases, complexity provides a more
accurate reflection of restrictions imposed by various research goals.
Finally, we }{outline several complexity metrics that can be used to
effectively evaluate mapping models.}

\end{abstract}


\begin{center}\rule{0.5\linewidth}{0.5pt}\end{center}

\hypertarget{h.u5g3gh6g6u25}{%
\section{Introduction}\label{h.u5g3gh6g6u25}}

{In recent decades, neuroscientists have witnessed a massive
increase in the }{amount of available data, as well as in the
computational power of the tools we can apply to the data. As a result, we can now leverage huge datasets to build powerful models of brain activity. In this era of new opportunities, it is important to be mindful of conceptual choices we make before modeling our data.}{~This paper discusses }{one such choice}{: the choice of }{a mapping model that relates}{~features of interest }{to}{~}{brain}{~responses. }

\begin{figure}
    \centering
    \includegraphics[width=\textwidth]{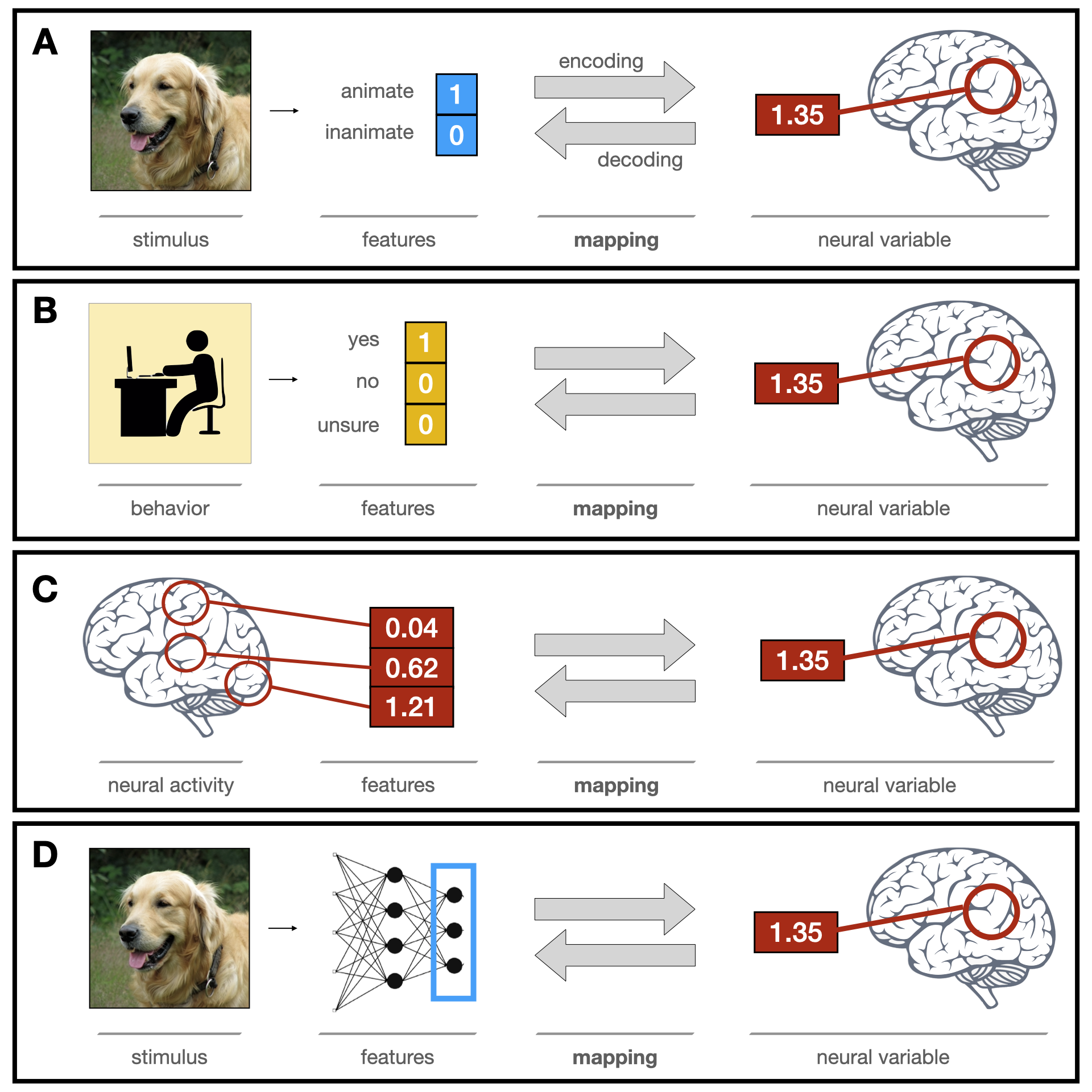}
    \caption{
    {The encoding/decoding modeling framework in cognitive
    neuroscience. }\textbf{(A)}{ Studies investigating the effect of external
    stimuli on brain activity start with the stimulus, extract its features
    of interest, and use a mapping model to establish the mapping between
    these features and a neural variable extracted from the data recorded
    during/after stimulus presentation. }\textbf{(B)}{ In other studies,
    }{researchers}{~extract features associated with participants' behavior
    and map those onto the neural variable recorded before/during this
    behavior. }\textbf{(C)}{ }{A}{nother class of studies}{~}{describes the mapping
    between activity in }{different}{~brain regions where neural variables
    serve as the features. }\textbf{(D)}{ In recent years, more and more studies
    replace ha}{nd-crafted features, like those shown in (A), with
    high-dimensional feature vectors derived from models of brain function,
    such as neural networks.}
    }
    \label{fig:encoding-decoding}
\end{figure}

When studying a brain circuit, area, or network, it is often useful to
formulate and test hypotheses about features that elicit a response
in the relevant neural
units\footnote{Note that the neural data being fitted is not necessarily the neural recording itself: researchers may choose to predict the average firing rate, power in a particular frequency band, or beta coefficients from the general linear model (GLM) of fMRI responses \citep{king_encoding_2020}.} 
(a single cell, a population of neurons, a brain area, etc.). {The
features can be stimulus-based (}\textbf{Figure \ref{fig:encoding-decoding}A}{), behavior-based (}\textbf{Figure \ref{fig:encoding-decoding}B}{), or }{based}{~on responses in other neural units within the same
brain or in a brain of another individual (}\textbf{Figure \ref{fig:encoding-decoding}C}{). The ways in
which the features are derived vary widely: common ~sources include
}{human annotations (e.g., ``faces'' and ``scenes'')}{,
empirical}{~measurements (e.g., behavioral or neural responses)}{, and
outputs of a computational model (e.g., a vector of responses to each
image in a layer of a deep neural network (DNN); }\textbf{Figure \ref{fig:encoding-decoding}D}{). }

{To relate a set of features to brain activity, we need to establish a
}{mapping}{~between them. }{The exact form of the mapping is typically
learned from the data, although the space of possible mappings is
defined in advance (e.g., linear functions). }

{Why is a mapping necessary? }{In principle, we could limit ourselves
to}{~mapping-free models, which use features of interest and a set of
fixed parameters to predict neural activity directly}{. For instance,
given some information about the stimulus, a mapping-free model would
predict the exact firing rate of a particular neuron or change in BOLD
activity in a given voxel.}{~However, mapping-free models in cognitive
neuroscience today }{are almost always infeasible}{. One limitation is a
mismatch between the granularity of our theoretic predictions and the
measurements to be modeled. For instance, we might want to test a model
of brain function that predicts an increased firing rate in a neuron in
response to a face but does not specify the exact amount of that
increase. }{The second limitation is the lack of }{a priori}{~knowledge
about }{functional differences between neural units that anatomy fails
to explain (e.g., which exact neurons will respond to faces in a given
individual). Thus, we might want to test our predictions against the
neural data without deciding }{a priori}{~which neural units (or
combinations whereof) might encode the features of interest. The third
limitation is the data that we use, which typically provide a noisy
}{and/or indirect measure of neural activity}{. The presence of noise
and/or an underdetermined linking function between our predictions and
our measurements means that we often want to incorporate some free
parameters to infer this information from the data. All in all, when
modeling neural data, }{some level of fitting is almost always required.
}

\begin{figure} 
    \centering
    \frame{\includegraphics[width=0.85\textwidth]{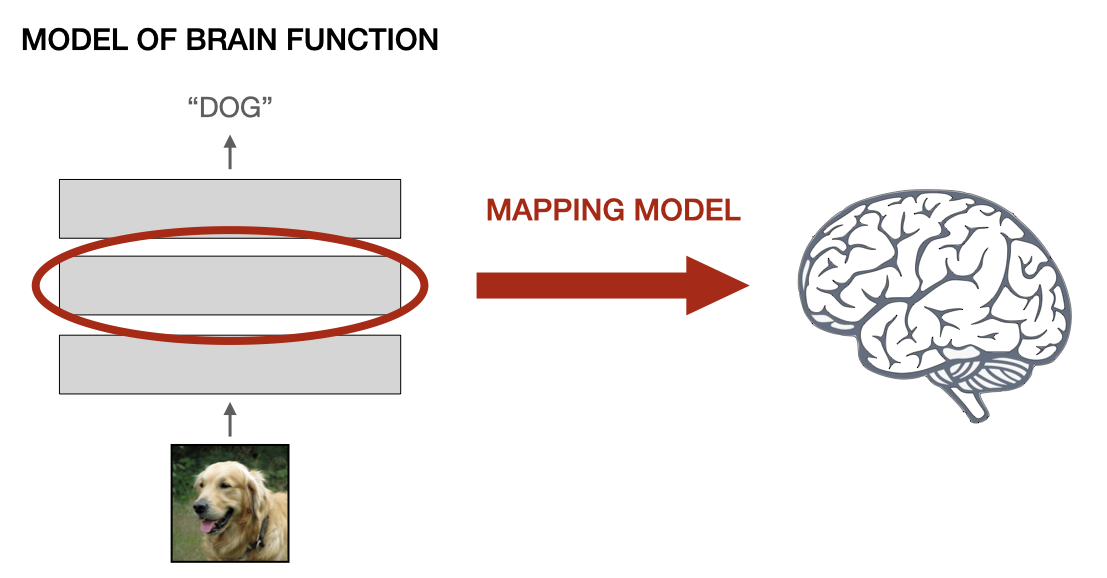}}
    \caption{The distinction between a model of brain function and a mapping model. A model of brain function aims to mimic the brain, but does not directly map onto neural data. Our focus is on mapping models, which link a feature set to a neural variable. The mapping model depicted here uses features derived from a model of brain function (like in \textbf{Figure \ref{fig:encoding-decoding}D}).}
    \label{fig:fig2-distinction}
\end{figure}

A mapping model is a model that relates features of interest and neural data\footnote{In this paper, we discuss both encoding mapping models, i.e. models that map from the features of interest to the neural variable, and decoding mapping models, i.e. models that map from the neural variable to the features of interest (\textbf{Figure \ref{fig:encoding-decoding}A}). Others have discussed the relative merits of the two approaches \citep{holdgraf_encoding_2017, king_encoding_2020, kriegeskorte_interpreting_2019, naselaris_encoding_2011}; our arguments in this paper apply to both mapping directions, unless specified otherwise. }.
Its main distinguishing feature is the presence of free parameters whose
values are determined in the process of training the model on neural
data. This makes mapping models meaningfully different from {models of
brain function}{, which aim to mimic neural computations but are not
trained on neural data (\textbf{Figure \ref{fig:fig2-distinction}}). In principle, a model of brain
function can also have its parameters trained or fine-tuned using neural
data
\citep{toneva_interpreting_2019}; in this case, the distinction between the two becomes blurred. In practice, however, the majority of studies today
separate these two steps: a model of brain function can be used to
derive features of interest, and a mapping model is then fitted to link
these features and neural data. }

{Mapping models have many properties that researchers need to take into
account, but the most common distinction is drawn between (a) a }{linear
mapping model}{~(such as linear regression) and (b) a
}{nonlinear}{~mapping model}{~(such as a neural network). }

\hypertarget{h.6rmw68a2hcd5}{%
\section{The controversy}\label{h.6rmw68a2hcd5}}

Today, the vast majority of cognitive neuroscience studies use linear
mapping models (such as linear regression). Linear mapping models are ubiquitous across domains (including vision, audition, and language) and recording techniques (from single unit recordings to non-invasive measures such as M/EEG and fMRI). Linear mapping models are often used
in conjunction with nonlinear models of brain function , a
common approach often referred to as a linearized model. In
linearized models, a predefined nonlinearity is applied to a stimulus and/or a neural variable to derive a set of features, which are then
linked to neural data using a linear mapping model. Linearized models
were first established in sensory neuroscience
\citep{aertsen_spectro-temporal_1981}{~and have become the predominant
approach, particularly in encoding models
}
\citep{naselaris_encoding_2011, van_gerven_primer_2017, wu_complete_2006}. 

{One of the major reasons that linear mapping models have become so
common is a practical one: they can be fitted using standard statistical
techniques that require little data and modest amounts of computation.
However, recent advances in machine learning (ML), coupled with an
increased availability of computing resources, have enabled
neuroscientists to explore a much broader range of modeling techniques
}\citep{bzdok_neuroimaging_2017}
{.}{~The concurrent increase in the size of available
datasets
}
\citep{chang_bold5000_2019, majaj_simple_2015, schoffelen_204-subject_2019}
{~}{has enabled researchers }{to train large-scale mapping models
}{without overfitting them.}{~}{As a result, a number of applied
neuroscience studies have leveraged the power of ML-based methods to
build flexible nonlinear mapping models and use them to identify neural
correlates of brain disorders
}
\citep{hasanzadeh_prediction_2019, kazemi_deep_2018, kim_deep_2016, leming_ensemble_2020}{}
{~and of behavioral traits
}
\citep{kumar_information_2019, morioka_nonlinear_2020, xiao_alternating_2019}.

{Although these computational advances have substantially diminished the
practical constraints on the mapping models (cf. \textbf{Section \ref{h.j4mss2teleyv}}),
many}{~researchers}{~still prefer to use linear mappings, arguing that
such mappings are preferable not only for practical, but also for
theoretical reasons. Some of the common theoretical arguments in favor
of linear mapping models are the following:}

\begin{enumerate}
\item
  {Li}{near }{mapping models }{facilitate a comparison of }{predictive
  accuracy}{~across feature sets
  }
  \citep{caucheteux_language_2020, jain_interpretable_2020, schrimpf_brain-score_2018, yamins_performance-optimized_2014}.
\item
  {L}{inear}{~mapping models estimate weights for individual features,
  }{making the mapping more
  }{interpretable}{~}
  \citep{anderson_predicting_2017, lee_masson_functional_2021, naselaris_encoding_2011, sudre_tracking_2012, haufe_interpretation_2014, kriegeskorte_interpreting_2019}.
\item
  {Linear mapping models are more }{biologically plausible}{: they
  approximate readout by a downstream area and can therefore indicate
  what information is available to the rest of the brain
  }
  \citep{kamitani_decoding_2005, kriegeskorte_pattern-information_2011}.
\end{enumerate}

{However, none of the three general desiderata that are relevant to
mapping model selection --- }{predictive accuracy}{, interpretability,
and biological plausibility --- align }{straightforwardly}{~with the
linear/nonlinear mapping model choice. }{In}{~the following section, we
critically review these desiderata and }{show that each of them, in
fact, can refer to several distinct research goals, each of which places
its own demands on the space of mappings to be considered, meaning that
the arguments above only apply in a subset of research scenarios and
cannot serve as general guidelines. }

\hypertarget{h.54ztrc2zgwlm}{%
\section{How do neuroscientists use mapping models?}\label{h.54ztrc2zgwlm}}

{T}{o }{choose}{~}{the best mapping }{model, we first need to specify
the goal that we are trying to achieve
}
\citep{kay_principles_2018, kording_appreciating_2020, kriegeskorte_pattern-information_2011}.
{~The goal, of
course, }{can be defined at various levels of granularity. Here, we
focus on specific goals a researcher might want to accomplish by
training a mapping model on their data (as opposed to high-level
research goals, such as curing Alzheimer's or building artificial
general intelligence). }

We categorize research goals under three general mapping model
desiderata\footnote{These desiderata are commonly used to argue in favor of or against certain mapping model choices. See, e.g., the online discussion of the proposal that served as a precursor to this paper: \url{https://openreview.net/forum?id=-o0dOwashib}} ---
predictive accuracy, interpretability, and biological plausibility.
We show that each desideratum, in fact, corresponds to several distinct
goals with their own mapping model requirements, and that the linearity
requirement only holds for some of the goals.

\hypertarget{h.x051f1yfrhdb}{%
\subsection{Predictive~accuracy}\label{h.x051f1yfrhdb}}

{In neuroscience, as in other fields, scientific progress is driven by
the generation of new hypotheses, followed by the testing of the
hypotheses' predictions against experimental data, and then by the
selection of the best (}{most accurate}{) hypothesis (or generation of
new hypotheses if none of the current hypotheses are good enough). In
the encoding/decoding framework, }{a hypothesis can be operationalized
as a set of features hand-crafted by the researchers}, (e.g.,
\citep{kanwisher_fusiform_1997}),
{derived from a model }{of brain function}
{(e.g.,
\citep{yamins_performance-optimized_2014})}
{~or obtained from behavioral ratings
}
(e.g., \citep{anderson_predicting_2017}).
%
{A common way to measure the predictive
accuracy of a set of features is to use a mapping model that will
estimate the best link between the features and the neural data. The
mapping --- fit on the training set --- can then be used to predict
responses in a held-out test set, after which we can evaluate those
predictions by, e.g., correlating them with the test data. These
correlations are often normalized by an estimate of the reliability of
the data (a ``ceiling'') to yield an estimate of }{explained
variance}
\citep{cadieu_deep_2014, dapello_simulating_2020, david_predicting_2005, geiger_wiring_2020, hsu_quantifying_2004, lage-castellanos_methods_2019, schoppe_measuring_2016, schrimpf_artificial_2020,talebi_natural_2012}.

{}

{This prediction-oriented framework can be used to achieve multiple
research goals, only some of which impose specific constraints on the
mapping model.}

{}

\subsubsection{Test feature relevance} \label{test-feature-relevance}
{One}{~research question a
neuroscientist might ask is ``do neural data $Y$ contain information about
features $X$?'' In}{ this }{scenario}{, }{the goal is to find
a}{~}{mapping model that allows us to achieve significant (above-chance)
predictive accuracy}{.}{~Above-chance model performance can have both
theoretical relevance (for instance, an encoding model showing that
features $X$ drive a certain brain region $Y$) and clinical relevance (for
instance, a decoding model that predicts disease severity $X$ based on
neuroimaging data $Y$). Formulated this way, the above-chance-accuracy
approach does not have to commit to the exact format of feature-brain
mappings --- }{all that matters is the mapping model's performance on
held-out data}{. For instance, a study that finds information about an
imagined visual scene ($X$) in primary visual cortex
}
\citep{naselaris_voxel-wise_2015}{}
{~could in principle provide a valuable
contribution to the field even if it used highly unconstrained nonlinear
mappings. Similarly, if researchers aim to predict whether a given case
of mild cognitive impairment will develop into Alzheimer's within the
next year, they do not have to pre-specify (or limit) the exact nature
of the mappings they will consider. All in all, for studies in this
category, the main objective is for the mapping model to achieve
significant predictive accuracy (or, for applied research, to reach a
certain accuracy threshold), and the space of possible mappings should
be large enough to allow it.}

\subsubsection{Compare feature sets} \label{compare-feature-sets}
{Predictive accuracy can be used to
}{compare competing feature sets (often extracted from different models
of brain function) }{~with the goal of selecting the one that best fits
neural data
}
\citep{schrimpf_integrative_2020}.
{This goal is achieved by fitting encoding mapping
models from several different feature sets to a pre-selected set of
neural responses and comparing the variance explained by each mapping. A
question that these studies tend to ask is: ``Which feature set provides
the most faithful reflection of the neural representational space?''
When performing a systematic comparison of feature sets, researchers
often choose to restrict themselves to linear mappings because }{a
powerful non-linear mapping model could inadvertently incorporate
transformations that reduce or erase the differences across feature
sets}{. For example, if the goal is to determine whether activity in
inferior temporal cortex is better predicted by an early or a late layer
of a convolutional neural network, we should use a mapping model with a
limited expressive power; otherwise, the mapping model will be able to
transform features from an early layer into features from a late layer,
eliminating meaningful differences between them. Thus, feature
comparison studies often benefit from linear mapping models. }

\subsubsection{Build maximally accurate models of brain
data}
{Finally, some researchers}{ are trying to build accurate
encoding models that, in essence, enable }{simulations of neuroscience
experiments} 
\citep{khosla2022high, ratan_murty_computational_2021}.
{This type of
modeling is especially important in cases when experimental data are
expensive or hard to obtain: with a high-accuracy model of brain
responses, a researcher can run thousands of experiments }\emph{in silico}{,
refine their hypothesis, and then test the critical predictions }\emph{in
vivo}{. Furthermore, these models may become an important component of
testing experimental replicability: if the same phenomenon is shown both
}\emph{in vivo}{ and }\emph{in silico}{, it is less likely to be a false positive.
In contrast to mapping models used for feature comparison (\textbf{Section \ref{compare-feature-sets}}),
models used for }\emph{in silico}{ experiments are only useful if they clear
a very high accuracy bar (ideally close to the noise ceiling of neural
data); otherwise, experiment simulations they produce would not be
trustworthy. The nature of the features used for these maximally
accurate models is also irrelevant: rather than testing the performance
of specific, pre-selected feature sets, the researcher can select any
set that achieves the desired results, even if these features end up
being raw picture pixels. The best way to build these }\emph{in silico}{ brains might be to train large powerful }{mapping models}{ on
large amounts of neural data. In this scenario, there is no theoretical
justification for a linear mapping constraint because, as in 
}\textbf{Section \ref{test-feature-relevance}}{,
the primary goal is maximizing predictive accuracy on held-out data.}

\hypertarget{h.woxacscg128f}{%
\subsection{Interpretability}\label{h.woxacscg128f}}

{Once }{we find}{~a mapping that achieves sufficiently high predictive
accuracy, we often want to interpret it. Which features contribute the
most to neural activity? Do neurons/electrodes/voxels respond to single
features or exhibit mixed selectivity? How does the mapping relate to
other models or theories of brain function?}

{The traditional view is that linear mappings are easier to interpret
than non-linear mappings
} \citep{naselaris_encoding_2011}.
{However, the goal of building }{interpretable models
is ultimately complicated by the fact that a clear-cut definition for
interpretability is lacking.}{~Below, we discuss three }{definitions of
interpretability, ranging from strictest to loosest, and show that they
provide different constraints on the mapping models.}{~Importantly, in
each of these cases, interpretability places restrictions not only on
the mapping model, but also on the features that can be used to yield
meaningful interpretations.}

\subsubsection{Examine individual weights in the mapping model}
{Traditionally, }{many }{cognitive neuroscientists have aimed to
interpret a neural signal by identifying a set of words to describe its
}{function }
(as in \citep{desimone_stimulus-selective_1984, kanwisher_fusiform_1997}).
{In this scenario, a useful
model of brain activity has features t}{hat can be described using one
or a few words (``faces'', ``vertical lines'', etc.) }{--- a property
that is often referred to as nameability ---}{~and a linear mapping
between these features and neural data}{. The dimensions of the neural
data are also nameable (e.g., a brain voxel with certain coordinates or
an electrode placed in a specific part of the brain), }{although}{~in
practice this criterion can be more loose. Under this setting, ~the
weights of a linear mapping model can be interpreted as a relative
measure of contribution of each input feature to each output feature
(this can be done regardless of mapping direction). . }{We consider this
to be the strictest definition of interpretability because it places the
strongest constraints on both the features (which have to be nameable)
and the space of possible mappings (which have to be linear). }

{Although this definition of interpretability is perhaps the most
intuitive, it suffers from two shortcomings. First, weight
interpretation is difficult in cases where the features (for an encoder)
or the brain recordings (for a decoder) suffer from multicollinearity
and/or when regularization is used to impose a prior }{on model weights
}
\citep{haufe_interpretation_2014, kriegeskorte_interpreting_2019}.
{These issues are
especially pervasive in the case of fMRI decoding models, where the
brain measurements from nearby voxels are almost always
correlated}{~}{and where researchers often use regression with L2
regularization (also called ridge regression) to deal with the large
number of voxels and noise in individual voxels. }{Second, }{feature
nameability }{may be an overly restrictive metric, as it limits our
understanding to a vocabulary that is heavily biased by \emph{a priori}
hypotheses and may not include words for the concepts we actually need
}
\citep{buzsaki_brain_2019}.
{For instance, }{recent work has shown that (a) neurons
typically described as ``face-responsive'' respond more strongly to
artificial images produced by DNNs than to natural images described by
the word ``face''
}
\citep{ponce_evolving_2019}.
{and (b) a neural-network-based linearized model of
activity in the fusiform face area predicts responses to faces better
than label-based models
}
\citep{ratan_murty_computational_2021},
{suggesting that simple verbal features cannot
provide a full account of neural activity.}{~To overcome the limitation
of using individual nameable features, many researchers have instead
started to use high-dimensional feature sets}{.}

\subsubsection{Test correspondences between representational
spaces}
{A looser notion of interpretability, which has become
popular in the last decade, relies on the use of high-dimensional
feature vectors that are linearly mapped, usually via encoding, to a
neural
}{variable}
\citep{kay_identifying_2008, yamins_performance-optimized_2014}.
{As mentioned in \textbf{Section \ref{h.6rmw68a2hcd5}},
this setup is commonly referred to as ``linearized'' models of brain
function}{. }{When using large-scale feature sets, }{we cannot always
interpret the weights of a linear mapping model in the same way as we
did with nameable }{features}{.}{~If individual features within a set
cannot be labeled and/or are derived via a sequence of nonlinear
operations (e.g., in the case of DNN layer activations), examining
individual features has a limited potential to inform our intuition
}
\citep{kay_principles_2018}.
{However, we can examine the feature set as a whole, to ask:
do features $X$, generated by a known process, accurately describe the
space of neural responses $Y$? Thus, the entire feature set becomes a new
unit of interpretation, and the linearity restriction is placed
primarily to limit the space of possible }{feature spac}{e
}{transformations}{.}{~For instance, the finding that convolutional
neural networks and the ventral visual stream produce representational
spaces that are similar up to a linear transformation
}
(e.g., \citep{yamins_performance-optimized_2014, caucheteux_language_2020, jain_incorporating_2018, schrimpf_brain-score_2018})
{allows us to infer that both processes are subject to similar optimization constraints
} \citep{richards_deep_2019}.
{That said, mapping models that relate two
representational spaces do not have to be linear, }{as long as they
correspond to a well-specified hypothesis about the relationship between
}{them; }{for instance, we might want to relate the intrinsic
dimensionality of the spaces being compared, an approach that is
inherently nonlinear
}
(\citep{chaudhuri_intrinsic_2019, gallego_cortical_2018}; for discussion, see \citep{jazayeri_interpreting_2021}).

\subsubsection{Describe the feature set as a whole}
{The loosest
definition of interpretability is }{the ability to name and/or
}{describe the set of features}{~that was used to train the mapping
model}{~(e.g., ``phonological features'')}{. }{In this scenario, we make
no assumptions about a particular representational geometry of these
features (such as linear separability).}{~}{The lack of specific
assumptions about the form of the feature-to-brain mapping means that
}{constraints on the mapping model}{~are not strictly necessary --- all
we need is }{an epistemologically satisfying description of the
features}{~that would apply regardless of which mapping is applied to
these features}{. If a mapping model that uses these features achieves
good predictivity, we can say that a given set of features is reflected
in the neural signal. }{Under this definition, any mapping model
(encoding or decoding) is interpretable as long as we can describe the
set of features that it }{uses}{.}

{}

\hypertarget{h.us64ws60pzah}{%
\subsection{Biological plausibility}\label{h.us64ws60pzah}}

{In addition to }{predictive accuracy}{~and interpretability-related
considerations, biological plausibility can also be a factor in deciding
on the space of acceptable feature-to-brain and brain-to-feature
mappings. Given that the mapping model}{~is commonly used to determine
which features are reflected in neural data}{, it is important to select
the space of mappings in a way that can lead to the selection of
biologically plausible feature sets. }

\subsubsection{Simulate linear readout}
{One of the main arguments in favor of
linear mappings is the claim that they approximate the linear readout
performed by a putative downstream brain area
} \citep{kamitani_decoding_2005, kriegeskorte_pattern-information_2011}.
{Under this view, the mapping model
approximates the transmission of the features to a hypothetical
information consumer. The linear readout requirement often serves as a
proxy for feature usability: if the features can be extracted with a
linear mapping model, it means that they require few additional
computations in order to be used downstream. Note that this argument
only applies to decoding mapping models, which match the direction of
the readout (from the brain area to a hypothetical consumer). }

{The ability to use features of interest in downstream computations is
indeed an important consideration.}{~}{However, there are reasons to be
cautious about the linear readout requirement. First, some models
operate on neural data that are collected from multiple recording sites
rather than a single neural population/region, which makes subsequent
linear readout biologically implausible. For instance, decoding models
that use whole-brain data, such as M/EEG, have no downstream region that
could `read out' information from all over the brain --- the only entity
performing readout is the observer. }{Second, linear readout }{might not
be an accurate characterization of the decoding mechanisms used by
downstream areas to extract information from the brain region of
interest. I}{n fact, unlike linear models that pool across all measured
neurons or voxels in the region of interest, readout in biological
neural systems is likely to be both sparse
}
(e.g., \citep{barak_sparseness_2013, barlow_trigger_1969, olshausen_sparse_2004, vinje_sparse_2000})
{and
}{nonlinear}
\citep{beniaguev_single_2021, ghazanfar_nonlinear_1997, gidon_dendritic_2020, jones_might_2021, shamir_nonlinear_2004}.
{Third, linear regression is a fairly arbitrary threshold
to draw for mechanistic plausibility. A linear mapping model can extract
many features from the data, some of which do not faithfully reflect the
underlying neural computations and could not possibly be read out by a
downstream neuron}{. For instance, fMRI signals from V1 contain
voxel-level biases that allow orientation decoding (such as radial
biases in the retinotopic map) that are distinct from
orientation-related neural computations (such as activity in
orientation-specific cortical columns), which results in a mismatch
between information used by the mapping model and information used by
actual neurons
}
\citep{ritchie_decoding_2019}.
{In sum, unconstrained linear mapping models (or linear
mapping models constrained by weight distribution among many features,
like ridge regression) may be both overly limiting because they do not
account for possible nonlinear computations and overly greedy because
they might leverage information in a way that real neurons do not.}

{Is there a better mapping model that accounts for possible nonlinear
computations during readout without being overly broad? }{One }{approach
is to introduce parsimony constraints on the feature spaces
}
\citep{kukacka_regularization_2017}. 
{For instance, introducing a}{~sparsity
constraint}{~(i.e., allowing the mapping model to access only a limited
number of neurons)}{~}{could increase the biological plausibility of
putative
readout} \citep{yoshida_natural_2020}.
{However, in the context of measurements that collapse
across large numbers of neurons (i.e., most measurements in cognitive
neuroscience), the sparsity constraint might be impossible to enforce,
as a single voxel or electrode already combines }{signal}{~from a large
number of neurons. More broadly, evaluating the biological plausibility
of decoding is difficult as readout might differ across brain regions
}{of interest
} \citep{anzellotti_beyond_2018},
{and the current understanding of the details of
readout mechanisms remains limited. Future progress in research on
readout mechanisms will be key to evaluate the different assumptions
about readout in a more principled manner.}

\subsubsection{Incorporate measurement-related considerations}
{W}{hen brain recordings are known to be nonlinear transformations of underlying neural activity}
{
{(e.g.,
fMRI, in which BOLD responses are related to neural responses via the
hemodynamic response function, or HRF \citep{friston_nonlinear_2000}
}},
{knowledge about the nonlinear relationship between the
neural responses and the measurements can (and often should) be
explicitly incorporated into the mapping. }{Failing to do so might
privilege feature sets that incorporate properties of the measurement
over feature sets that more accurately reflect the neural
representations encoded in a brain region but happen to be nonlinearly
related to the measured signal.}{~}

{The issue of nonlinear properties of the measurement is prominent in
fMRI analyses}{.}{~The traditional approach to fMRI data fitting is a
linearized encoding model: the predictor variable is convolved with the
HRF, and the resulting predictor is then linearly fitted to the data.
However, these models typically assume a fixed HRF across voxels
}
\citep{friston_nonlinear_2000}
{and/or conditions
}
\citep{pedregosa_data-driven_2015}
{which is not biologically plausible}
\citep{ekstrom_regional_2021, handwerker_variation_2004}.
{More flexible finite impulse response models}
(FIR; \citep{dale_optimal_1999, glover_deconvolution_1999})
{require fitting a large number of
parameters and are computationally brittle. Thus, instead of sticking
with linearized approaches, some researchers are suggesting to model the
HRF shape explicitly}{~within a family of nonlinear functions motivated
by physiological data
}
(for instance, \citep{lindquist_validity_2007, shain_fmri_2020, shain_cdrnn_2021}).
{By using a constrained space of nonlinear mappings (rather
than an unconstrained space of linear mappings, as in FIR), one can
estimate the veridical shape of the HRFs using a relatively small number
of parameters. In principle, the nonlinear modeling approach can be
applied to both encoding and decoding fMRI studies, although in practice
the encoding direction is more feasible.}

{M/EEG analyses face similar issues. A common approach (for both
encoding and decoding) is to fit a linear mapping between the predictors
and the M/EEG measurements or their derivative (e.g., power in a
particular frequency band). However, the predictor features taken from
the best linear model might inadvertently incorporate the nonlinear
mapping between the feature and the captured response. Further, for
M/EEG, the recorded signal is a combination of both inhibitory and
excitatory signals; thus, treating it as a straightforward linear
combination is not always possible
}
\citep{hansen_meg_2010}.
{Thus, linear mapping models often overlook the complexities of neuroimaging signals, sacrificing biological
plausibility as a result.}

{To summarize \textbf{Section \ref{h.54ztrc2zgwlm}}, different research goals place different
constraints on the mapping model. A particular goal might require using a generic
linear mapping model, adding additional restrictions to that model,
using a particular class of nonlinear models, or imposing no \emph{a priori}
restrictions on the mapping space. }

\hypertarget{h.j4mss2teleyv}{%
\section{Practical considerations}\label{h.j4mss2teleyv}}
{The criteria outlined above are primarily based on theoretical
considerations: which mapping model has the properties that allow us to
achieve a particular goal? However, another important consideration is
practical feasibility: do we have enough data to accurately estimate the
mapping? Will the noise in our data lead certain mapping models to
fail?}

{Determining how much data is required for fitting a particular mapping
model has critical implications for experimental design (the number of
trials/data points per participant, the number of repetitions per
stimulus, etc.). In general, the fewer constraints are placed on the
mapping model, the more data will be needed to converge on a good
mapping. This relationship can be estimated empirically using standard
validation methods by, for instance, taking a large dataset and
evaluating the mapping model's predictive accuracy on left-out test data
while gradually increasing the size of the training dataset. However,
few studies report such analyses (and in some cases, large-enough
datasets may still be lacking). One exception is a line of fMRI studies
that aim to determine the best mapping model for linking interregional
functional correlations and behavioral/demographic traits. The results
of these studies are mixed: some report a marked advantage of nonlinear
mapping models over linear ones
} \citep{bertolero_deep_2020}
{whereas others report that linear mapping models
perform equally well even when the training set includes several
thousand brain images
}
\citep{he_deep_2020, schulz_different_2020}.
{Thus, the field would greatly
benefit from further systematic examinations of the influence of dataset
size (}{and other experimental design properties}{) on the performance
of a particular mapping model type.}

{Even with large amounts of data, certain measurement properties might
force us to use a particular mapping class. For instance, 
}
\citet{nozari_is_2020}
{show
that the relationship between activity in different brain regions during
rest, as captured by fMRI, is best modeled with linear mappings and
suggest that fMRI's inevitable spatiotemporal signal averaging might be
to blame
}
(although see \citep{anzellotti_measuring_2017} for contrary evidence).
{In sum, even after establishing theoretical
desiderata for the mapping model, we need to conduct rigorous empirical
tests to determine which mapping model class will achieve good
predictive accuracy on test data given the measurement technique, the
amount and quality of available data, and other practical
considerations.}

\hypertarget{h.lcdrvwt16666}{%
\section{Going forward: evaluating mapping model complexity}\label{h.lcdrvwt16666}}

{Instead of focusing exclusively on the linear/nonlinear dichotomy, we
propose to reframe the choice of mapping model in the context of a
broader notion of model complexity. Complexity lies at the heart of most
desiderata discussed above. Arbitrarily complex models make predictive
accuracy comparisons across feature sets more difficult; they can be
harder to interpret; and }{they are }{perhaps}{~less likely to match
computations in biological circuits}{. }{Thus, we suggest replacing the
linear/nonlinear dichotomy with a framework that takes into account the
complexity of the mapping model. }

\hypertarget{h.vysnu1lu8ny8}{%
\subsection{The role of complexity in selecting a mapping model}\label{h.vysnu1lu8ny8}}

As we saw in \textbf{Section \ref{h.54ztrc2zgwlm}}, specific research goals impose different
constraints on the mapping model. Further, these constraints are often
more graded than the linear/nonlinear distinction and can instead be
seen as restrictions on model complexity: 

\begin{itemize}
\item
  {Interpreting individual features is easier when the mapping is not
  only linear, but also sparse, so that each neuron can be described
  with only a few features. Reframing the mapping model choice in terms
  of complexity allows us to pick out simple mappings }{within}{~the
  class of linear mapping models, thus facilitating interpretation. }
\item
  {Satisfying biological constraints, such as accounting for
  physiological properties of the measurement or simulating neural
  readout, may require a certain degree of nonlinearity but these
  nonlinearities are often well-defined and can keep overall model
  complexity relatively low. }
\item
  {Testing whether a feature set accurately captures the
  representational space of neural responses may require the mapping to
  preserve certain properties of that space. Here, the complexity of the
  mapping model depends primarily on the hypothesis being tested.}
\item
  {Comparing and/or interpreting feature sets is possible even when the
  mapping is nonlinear, as long as we can compare the mappings using a
  metric that incorporates both predictive accuracy and model
  complexity. }
\item
  {Decoding features from neural data and building accurate encoding
  models of the brain does not require placing any theory-based
  restrictions on the mapping model (although such restrictions might
  improve performance in practice).}
\end{itemize}

\begin{figure}
    \centering
    \includegraphics[width=\textwidth]{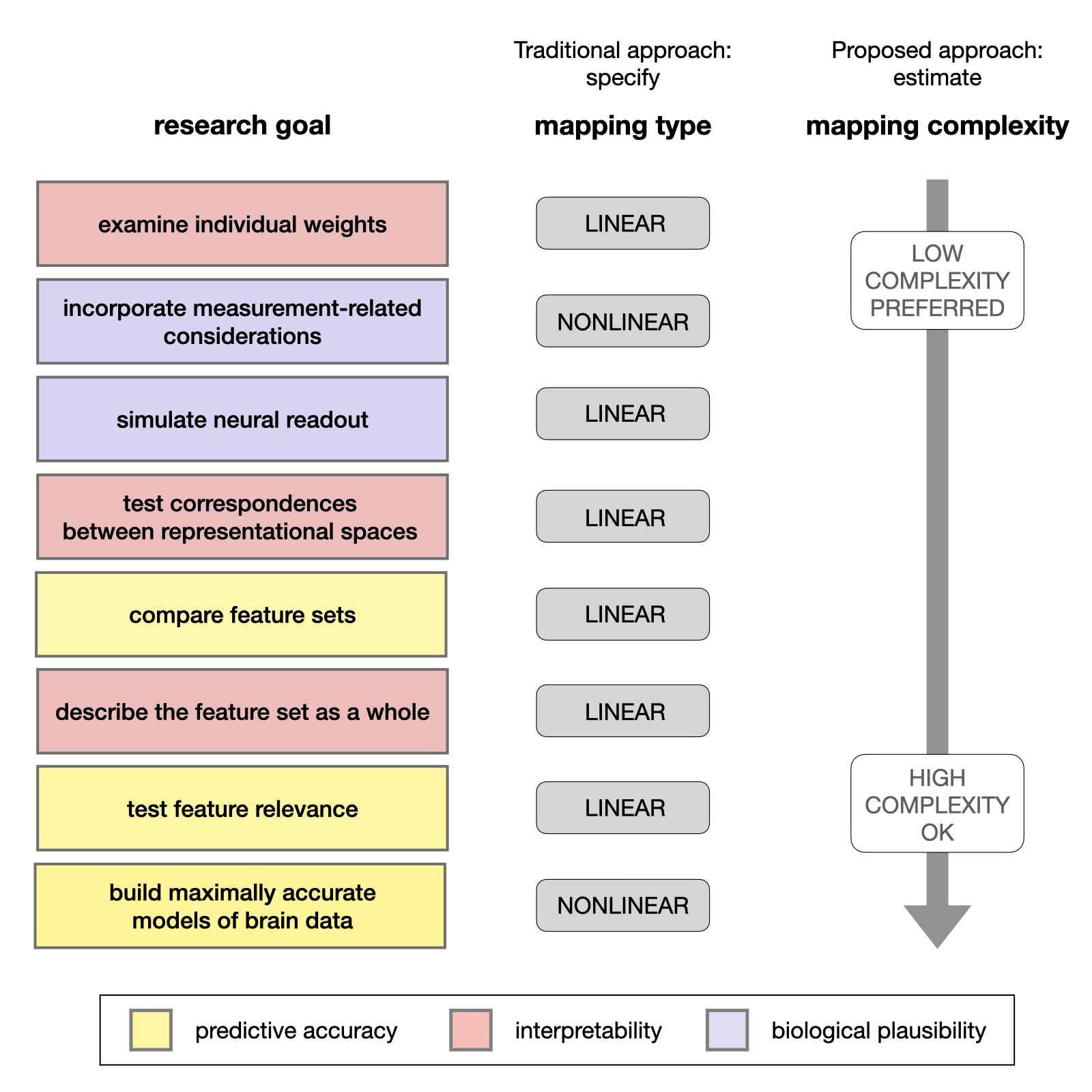}
    \caption{{Different research }{goals}{~are currently being
    collapsed into the ``linear/nonlinear'' dichotomy but in fact correspond to different degrees of mapping model complexity}{.}{~Note that the exact ordering of research goals along the complexity continuum is approximate and shown primarily for illustration purposes.}}
    \label{fig:research-goals-model-complexity}
\end{figure}

\textbf{Figure \ref{fig:research-goals-model-complexity}} shows the research goals discussed above together with the
mapping model types that are traditionally used to achieve these goals,
as well as our proposal to shift from the linear/nonlinear dichotomy to
explicit estimates of ~model complexity. Note that this diagram depicts
theoretical, \emph{a priori} criteria for restricting mapping model
complexity; practical considerations might impose additional constraints
to achieve better predictivity (see \textbf{Section \ref{h.j4mss2teleyv}}). 

\hypertarget{h.n7avh2vzlypy}{%
\subsection{Complexity measures}\label{h.n7avh2vzlypy}}

{How can we estimate the complexity of mapping models? To date, many
studies have focused primarily on a binary distinction in which linear
models are ``simple'' and nonlinear models are ``complex''. However, as
discussed above, this distinction is overly simplistic. Here, we review
several measures of mapping model complexity that are commonly used in
the ML literature and may serve as an alternative to the
linear/nonlinear dichotomy (see }\textbf{Table \ref{table:complexity}}{ for a side-by-side
comparison). }

\subsubsection{Number of free parameters}
{A common approach to measuring
model complexity is by considering the number of free parameters in the
model. In this approach, each model class encounters a penalty that
corresponds to its number of parameters, such that classes with more
parameters have a larger penalty. In order to justify the use of
additional parameters, a model needs to achieve a substantial
performance improvement compared to models with fewer parameters. }{This
}{tradeoff is often implemented using Akaike's Information Criterion
(AIC) or the Bayesian Information Criterion (BIC), which reward models
for good predictive performance but penalize them for the number of
parameters. Although simple to estimate, this complexity measure often
fails to capture distinctions that seem intuitively important. For
instance, a linear and a nonlinear model with the same number of
parameters would have equal complexity in this view, even though the
latter often has a greater expressive power. Another example is a sparse
mapping model that allows non-zero weights only for a few features vs. a
dense model that places non-zero weights on, say, 500 features: if the
initial feature vector size is the same, then these models will have the
same number of parameters and therefore equal complexity under this
measure.}

\subsubsection{Minimum description length}
{Another common approach to measuring model complexity is based on minimum description length}
(MDL; \citep{rissanen_modeling_1978}).
{This approach typically assumes an encoding
function, i.e., a formal description language, over a class of models,
and the complexity of each model within the class is determined by the
length of the model's encoding. The encoding function essentially serves
as a prior over the model class: more probable mapping models would be
assigned shorter descriptions
}
{{(see \citep{diedrichsen_representational_2017, wu_complete_2006}{}
for a discussion of
the relationship between priors and regularization constraints)}}{. The
MDL approach can overcome some of the limitations of complexity measures
based solely on the number of free parameters by exploiting correlations
between parameters to achieve a shorter description length. For
instance, under this scheme, sparse models can have a shorter
description length and would therefore be considered less complex.
However, the main limitation of }{an}{~MDL-based metric is that it
requires specifying a mapping model class, as well as an encoding scheme
for mapping models within that class}{.}Thus, if there is no natural
prior over the set of mappings we wish to compare, an architecture-free
complexity measure may be
preferred\footnote{For example, certain informational measures \citep[e.g.,][]{bialek_predictability_2001, gilad-bachrach_information_2003} can be used to measure the complexity of the statistical relationship between the inputs and outputs of the mapping model (e.g., features and predicted neural data) regardless of a particular architecture or model class and, in some cases, may also capture the complexity of non-parametric generative models.}.

\subsubsection{Sample complexity}{Finally, a more
}{practice-oriented}{~metric is }{sample complexity
}
\citep{kearns_introduction_1994}.
{Loosely speaking, the sample complexity of a
model class is a function that determines the minimal number of training
samples needed to achieve desired model performance. It is not always
straightforward to compute this function } \emph{a priori}{; however, it can
be assessed empirically by computing learning curves, i.e., the achieved
level of predictive accuracy on a test set as a function of the number
of training samples. Estimates of sample complexity are vital for
understanding whether a given model failed because the underlying
hypothesis was wrong or because the dataset was too small to achieve a
proper }{fitting}{.}

\begin{table}
\begin{tabular}[]{@{}p{.25\textwidth}p{.33\textwidth}p{.33\textwidth}@{}}
\toprule
{\bf Complexity metric} & {\bf Benefits} & {\bf Limitations} \\
\midrule
{Number of parameters} & {Straightforward estimation} & {Does not
always capture relevant complexity distinctions (e.g., sparsity
or model architecture}{)} \\
\midrule
{Minimum description length \newline (MDL)} & {Better at capturing relevant
distinctions (e.g., sparsity) }{by accounting for correlations
between parameters} & {Requires a prior distribution over the mapping
model class, which is not always easy to specify} \\
\midrule
{Sample complexity} & {Comparable across model architectures; immediate
practical application} & {Does not necessarily capture the expressive
power of the model class} \\
\bottomrule
\end{tabular}
\caption{Benefits and limitations of several mapping model complexity measures.}
\label{table:complexity}
\end{table}

{In summary, instead of defaulting to linear models, we propose to
incorporate a measure of model complexity into the general evaluation
framework of encoding/decoding models. This measure can be used in
different ways depending on the research goal. }{For}{~instance, for
}{feature comparison, if two feature sets produce equally accurate
mapping models, the feature set corresponding to a }{simpler}{~mapping
model (as measured with minimum description length) may represent a
better fit to neural data. }{For estimates of potential downstream
readout, instead of limiting ourselves to linear functions, we can
consider a range of possible mappings, where simpler mappings }{reflect
a higher probability that these features are used downstream}{. Thus,
measuring model complexity can serve as a powerful tool for mapping
model evaluation and selection.}

\hypertarget{h.vnuan6wdd3m0}{%
\section{Conclusion}\label{h.vnuan6wdd3m0}}

{The encoding/decoding framework in contemporary cognitive neuroscience
has provided many valuable insights. However, in some cases, the field
has been held back by its excessive reliance on linear mappings between
features and brain activity. Here, we have described various research
goals that are typically considered when specifying a mapping model.
Contrary to popular belief, few of these goals require the use of linear
mapping models. Instead, some do not require placing }{any}{~constraints
on the mapping model, some require placing specific }{nonlinear
}{constraints, and some use linearity simply as a proxy for reducing
model complexity. We therefore }{suggest to
explicitly}{~}{include}{~measures of model complexity when selecting and
evaluating mapping models. An increased focus on mapping model
complexity could }{help the field}{~discover a richer space of accurate,
simple, biologically plausible predictors of neural }{activity}{, thus
advancing our overall understanding of brain function.}

\hypertarget{h.b17xloclsav9}{%
\section{Acknowledgements}\label{h.b17xloclsav9}}

{This paper is part of the Generative Adversarial Collaboration (GAC)
initiative organized by the Computational Cognitive Conference board. We
thank the GAC organizers, especially Megan Peters and Gunnar Blohm, for
their invaluable help with this initiative. Many of the ideas discussed
in this work arose during the GAC workshop in October 2020
(
\href{https://www.google.com/url?q=https://www.youtube.com/watch?v\%3DUI5KclR71IE\%26list\%3DPLNWftEg2R4s5iObSUvPXhnDJvyNbs4PnM\%26index\%3D2\&sa=D\&source=editors\&ust=1660235851855443\&usg=AOvVaw1nwNVir3MrUFV8jR-KFntd}{https://www.youtube.com/watch?v=UI5KclR71IE\&list=PLNWftEg2R4s5iObSUvPXhnDJvyNbs4PnM\&index=2}).
We thank the invited workshop speakers --- Kohitij Kar, Mariya Toneva,
Laura Gwilliams, Jean-Rémi King, Martin Hebart, and Anna Schapiro---as
well as workshop participants for their ideas, comments, questions, and suggestions. ~We also thank the reviewers who provided comments on our
GAC proposal (the reviews are available at
}{\href{https://www.google.com/url?q=https://openreview.net/forum?id\%3D-o0dOwashib\&sa=D\&source=editors\&ust=1660235851855892\&usg=AOvVaw0t3psH5jHLli7MjlXhfW0-}{https://openreview.net/forum?id=-o0dOwashib}}) and on earlier versions of this paper.


\bibliography{anya}





\end{document}